\documentclass[twocolumn,showpacs,preprintnumbers,amsmath,amssymb,superscriptaddress]{revtex4}

\usepackage{graphicx}
\begin{document}
\bibliographystyle{prsty}
\title{Hybridization between the conduction band and $3d$ orbitals in the oxide-based diluted magnetic semiconductor In$_{2-x}$V$_x$O$_3$}

\author{M.~Kobayashi}
\affiliation{Department of Physics, University of Tokyo, 
7-3-1 Hongo, Bunkyo-ku, Tokyo 113-0033, Japan}
\author{Y.~Ishida}
\altaffiliation{Present address: RIKEN, SPring-8 Center, Sayo-cho, Hyogo 679-5148, Japan}
\affiliation{Department of Physics, University of Tokyo, 
7-3-1 Hongo, Bunkyo-ku, Tokyo 113-0033, Japan}
\author{J.~I.~Hwang}
\affiliation{Department of Physics, University of Tokyo, 
7-3-1 Hongo, Bunkyo-ku, Tokyo 113-0033, Japan}
\author{G.~S.~Song}
\affiliation{Department of Physics, University of Tokyo, 
7-3-1 Hongo, Bunkyo-ku, Tokyo 113-0033, Japan}
\author{M.~Takizawa}
\affiliation{Department of Physics, University of Tokyo, 
7-3-1 Hongo, Bunkyo-ku, Tokyo 113-0033, Japan}
\author{A.~Fujimori}
\affiliation{Department of Physics, University of Tokyo, 
7-3-1 Hongo, Bunkyo-ku, Tokyo 113-0033, Japan}
\author{Y.~Takeda}
\affiliation{Synchrotron Radiation Research Unit, 
Japan Atomic Energy Agency, Sayo-gun, Hyogo 679-5148, Japan}
\author{T.~Ohkochi}
\affiliation{Synchrotron Radiation Research Unit, 
Japan Atomic Energy Agency, Sayo-gun, Hyogo 679-5148, Japan}
\author{T.~Okane}
\affiliation{Synchrotron Radiation Research Unit, 
Japan Atomic Energy Agency, Sayo-gun, Hyogo 679-5148, Japan}
\author{Y.~Saitoh}
\affiliation{Synchrotron Radiation Research Unit, 
Japan Atomic Energy Agency, Sayo-gun, Hyogo 679-5148, Japan}
\author{H.~Yamagami}
\affiliation{Synchrotron Radiation Research Unit, 
Japan Atomic Energy Agency, Sayo-gun, Hyogo 679-5148, Japan}
\affiliation{Department of Physics, Faculty of Science, 
Kyoto Sangyo University, Kyoto 603-8555, Japan}
\author{Amita~Gupta}
\affiliation{Department of Materials Science-Tmfy-MSE, Royal Institute of Technology, 
Stockholm SE 10044, Sweden}
%\affiliation{Materials Science Division, BARC, Mumbai-40085, India}
\author{H.~T.~Cao}
\affiliation{Department of Materials Science-Tmfy-MSE, Royal Institute of Technology, 
Stockholm SE 10044, Sweden}
\author{K.~V.~Rao}
\affiliation{Department of Materials Science-Tmfy-MSE, Royal Institute of Technology, 
Stockholm SE 10044, Sweden}
\date{\today}

\begin{abstract}
The electronic structure of In$_{2-x}$V$_x$O$_3$ ($x=0.08$) has been investigated using photoemission spectroscopy (PES) and x-ray absorption spectroscopy (XAS). The V $2p$ core-level PES and XAS spectra revealed trivalent electronic state of the V ion, consistent with the substitution of the V ion for the In site. The V $3d$ partial density of states obtained by the resonant PES technique showed a sharp peak above the O $2p$ band. 
While the O $1s$ XAS spectrum of In$_{2-x}$V$_x$O$_3$ was similar to that of In$_2$O$_3$, there were differences in the In $3p$ and $3d$ XAS spectra between V-doped and pure In$_2$O$_3$. The observations give clear evidence for hybridization between the In conduction band and the V $3d$ orbitals in In$_{2-x}$V$_x$O$_3$.

\end{abstract}

%\pacs{71.28.+d, 71.30.+h, 79.60.Dp, 73.61.-r}

\maketitle
%1. INTRODUCTION

%1-1. Diluted magnetic semiconductor
Diluted magnetic semiconductors (DMS's) have been studied intensively since the discovery of ferromagnetism in the III-V DMS Ga$_{1-x}$Mn$_x$As \cite{APL_96_Ohno, PRB_98_Matsukura} because the ferromagnetic interaction between the Mn ions mediated by hole carriers enables us to manipulate both the charge and spin degrees of freedom of electrons \cite{Science_98_Ohno}. 
%DMS's are considered to become key materials for spin electronics or spintronics. 
Ferromagnetic DMS's having Curie temperature ($T_\mathrm{C}$) above room temperature have been strongly desired for realistic spintronic applications. 
%1-2. Oxide-based DMS
Ever since the theoretical prediction by Dietl {\it et al.} \cite{Science_00_Dietl} that Mn-doped GaN and ZnO with high hole concentrations should show ferromagnetism above room temperature, wide-gap semiconductors have become promising host materials for high-$T_\mathrm{C}$ DMS. In fact, there have been many reports on room temperature ferromagnetism in oxide-based DMS's such as Co-doped TiO$_2$ \cite{Science_01_Matsumoto}, Co-doped SnO$_2$ \cite{PRL_03_Ogale}, and Mn-doped ZnO \cite{NatMater_03_Sharma}. 
%1-3. In$_2$O$_3$-based DMS

Recently, In$_2$O$_3$-based DMS's have attracted much attention because of the reports of room temperature ferromagnetism in not only light transition-metal (TM) but also heavy TM doped In$_2$O$_3$ \cite{APL_04_Philip, APL_05_Yoo, APL_05_Hong, APL_06a_Peleckis, JPCM_06_Hong, APL_06b_Peleckis, JAP_07_Gupta, APL_07_Jayakumar, APL_05_He, SSC_06_Kim, NatMater_06_Philip, PRB_06_Yu, PRB_07_Stankiewicz} and of the potential of the host material In$_{2}$O$_{3}$ for applications. 
Here, In$_{2}$O$_{3}$ has a band gap of $\sim 3.5$ eV, is an $n$-type semiconductor, and crystallizes in the cubic bixbyite structure, where In atoms are coordinated by six oxygens forming octahedral ($O_h$) and orthorhombic ($D_{2h}$) octahedra and neighboring octahedra are shared with their corners and edges \cite{PRB_97_Tanaka, PRB_04_Marsella}, and Sn-doped In$_{2}$O$_{3}$ (ITO) has been famous for its  high electrical conductivity and transparency \cite{JAP_99_Kim, TSF_02_Granqvist}. 
Magnetic force microscopy observations of In$_{2-x}$Cr$_x$O$_{3-\delta}$ and In$_{2-x}$Ni$_x$O$_3$ have demonstrated that the topological undulations correspond to the strength of magnetic response, indicating a uniform distribution of magnetic domains in these materials \cite{NatMater_06_Philip, JPCM_06_Hong}. 
Reports of In$_2$O$_3$-based DMS's have suggested relationship between the electrical conductivity and ferromagnetism through measurements of, e.g., anomalous Hall effects \cite{APL_05_He, SSC_06_Kim, NatMater_06_Philip, PRB_06_Yu, PRB_07_Stankiewicz}. 
As In$_2$O$_3$-based DMS's are candidates for room-temperature ferromagnetic DMS's, the knowledge of their electronic structure is necessary to understand of the origin of the ferromagnetism.

%1-4. PES and XAS
Photoemission spectroscopy (PES) and x-ray absorption spectroscopy (XAS) are powerful tools to investigate electronic structure of materials. 
XAS, which means photon absorption from a core-level electron into unoccupied states, is an element specific technique to study the electronic structure. 
$2p \to 3d$ resonant photoemission spectroscopy (RPES) enables us to extract the $3d$ partial density of states (PDOS) in the valence band. 
%In $n$-type DMS
%Since the mechanism of the ferromagnetism is strongly related to the electronic structure of the $3d$ transition-metal ion and to the positions of the $3d$ PDOS in the valence band relative to the Fermi level ($E_\mathrm{F}$) \cite{SST_02_Sato, PRB_01_Okabayashi}, it is considered that RPES and XAS measurements would be beneficial to investigate DMS \cite{PRB_99_Okabayashi}. 
In this work, we have performed PES and XAS measurements on In$_{2-x}$V$_x$O$_3$ (IVO) thin films in order to obtain a fundamental understanding of the electronic structures. The valence state of the doped V ion has been determined by core-level x-ray photoemission spectroscopy (XPS) and XAS. Effects of doping on the electronic structure of the host material have been approached by V $2p \to 3d$ RPES and XAS at the O and In edges.

%2. EXPERIMENTAL
%2-1. Sample preparation
Each of In$_2$O$_3$ and In$_{2-x}$V$_x$O$_3$ ($x=0.08$) thin films was highly oriented normal to the plane of a sapphire(0001) substrate by the pulsed laser deposition technique. During the deposition, the substrate temperature was kept at $\sim 400$ $^{\circ}$C. The total thickness of the deposited layer was $\sim 500$ nm. X-ray diffraction confirmed that the thin film had the cubic bixbyite structure and no secondary phase was observed. Details of the sample fabrication are given in Ref.~\cite{JAP_07_Gupta}. The V concentration $x$ in the thin films was estimated from the intensity of the V $2p$ core-level PES spectrum. Ferromagnetism above room temperature was confirmed by magnetization measurements using a SQUID magnetometer (Quantum Design, Co. Ltd.).

%2-2. Experimental condition for PES measurements
RPES and XAS measurements were performed at the soft x-ray beam line BL23SU of SPring-8 \cite{NIMPRSA_01_Saitoh, AIP_04_Okamoto}. The monochromator resolution was $E/{\Delta}E > 10,000$. XAS signals were measured by the total electron yield method. The background of the XAS spectra was assumed to be a hyperbolic tangent function. 
The RPES and XPS measurements were performed in a vacuum below $1.0{\times}10^{-7}$ Pa using Gammadata Scienta SES-2000 and SES-100 hemispherical analyzers, respectively. A Mg-$K\alpha$ x-ray source ($h{\nu}=1253.6$ eV) was employed for the XPS measurements. 
In the PES measurements, photoelectrons were collected in the angle integrated mode at room temperature. The total resolution of the RPES and XPS measurements including temperature broadening were $\sim 300$ and $\sim 800$ meV, respectively. Sample surface was cleaned by cycles of Ar$^+$-ion sputtering. Cleanliness of the sample surface was checked by the absence of a high binding-energy shoulder in the O $1s$ spectrum and C $1s$ contamination by XPS. 
The position of the Fermi level ($E_\mathrm{F}$) was determined by measuring PES spectra of evaporated gold which was electrically in contact with the samples.

\begin{figure}[b!]
\begin{center}
\includegraphics[width=8.5cm]{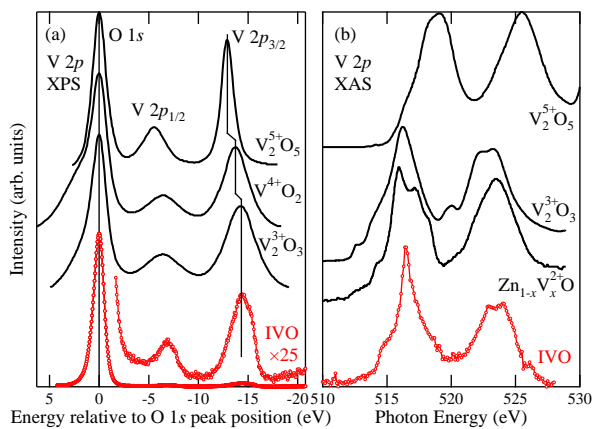}
\caption{V $2p$ core-level spectra of In$_{2-x}$V$_{x}$O$_3$ (IVO) with $x=0.08$. 
(a) V $2p$ XPS spectra compared with those of other vanadium oxides \cite{PRB_79_Sawatzky}. 
(b) V $2p$ XAS spectra compared with those of other vanadium oxides Zn$_{1-x}$V$_x$O \cite{APL_07_Ishida}, V$_2$O$_3$ \cite{DrThesis_Park}, and V$_2$O$_5$ \cite{JAP_07_Gloskovskii}. 
}
\label{V2pSpectra}
\end{center}
\end{figure}

%3. RESULTS AND DISCUSSION
%3-1. Valency of the V ion
First, we discuss about the valence state of the V ions in IVO. 
Figure~\ref{V2pSpectra} shows the V $2p$ core-level spectra of IVO and various vanadium oxides. 
The binding energy ($E_B$) of a core-level peak position is related to its charge state and, in general, becomes larger with increasing valency. In fact, it has been reported that the $E_B$ of the V $2p_{3/2}$ peak increases with valency of V as shown in Fig.~\ref{V2pSpectra}(a) \cite{PRB_79_Sawatzky}. 
By comparing the peak position of the V $2p_{3/2}$ core level of IVO with those of the other vanadium oxides, the valence state of V in In$_2$O$_3$ is found to be trivalent V$^{3+}$ ($d^2$). 
Figure~\ref{V2pSpectra}(b) shows the V $2p$ XAS spectra of IVO and several vanadium oxides. The line shape of the V $2p$ spectrum of IVO is similar to that of V$_2$O$_3$ (V$^{3+}$) \cite{DrThesis_Park} rather than that of Zn$_{1-x}$V$_x$O (V$^{2+}$) \cite{APL_07_Ishida} and of V$_2$O$_5$ (V$^{5+}$) \cite{JAP_07_Gloskovskii}, consistent with the observation of the V $2p$ XPS. The results suggest that the electronic structure of the doped V ion in IVO is close to that of the V ion in V$_2$O$_3$, i.e., V is in the trivalent V$^{3+}$ state octahedrally coordinated by oxygens. 
Therefore, it is likely that the doped V ions are substituted for the In sites.

\begin{figure}[b!]
\begin{center}
\includegraphics[width=8.5cm]{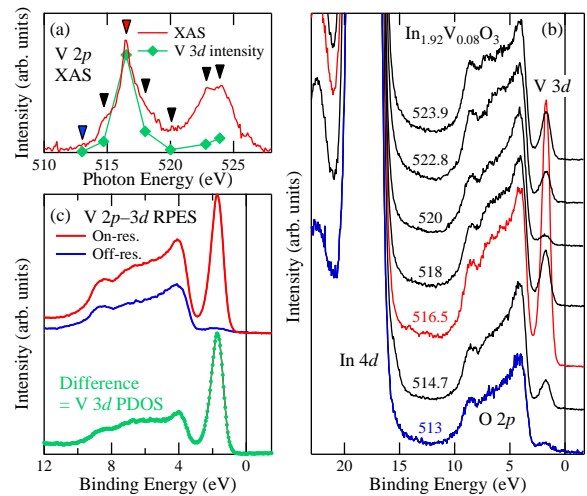}
\caption{V $2p \to 3d$ resonant photoemission spectra in the valence band of of In$_{2-x}$V$_x$O$_3$ ($x=0.08$). 
(a) V $2p$ XAS spectrum and the intensity of the V $3d$ peak at $E_B=1.7$ eV as functions of photon energy. 
(b) A series of spectra measured at $h\nu$'s denoted by triangles in (a). 
(c) Top: On- ($h\nu = 516.5$ eV) and off-resonance ($h\nu = 513$ eV) spectra. Bottom: On- and off-resonance difference spectrum representing the V $3d$ partial density of states (PDOS). 
}
\label{V3dPDOS}
\end{center}
\end{figure}

%3-2. V 3d PDOS
In order to obtain an understanding of the electronic structure of the V ion in the valence band, we measured V $2p \to 3d$ resonant photoemission spectra of IVO. Figure~\ref{V3dPDOS}(b) shows the valence-band PES spectra of IVO taken at various photon energies in the V $2p \to 3d$ core-excitation region. 
Energy difference between the top of the O $2p$ bands and $E_\mathrm{F}$ is $\sim 3.0$ eV, comparable to the band gap of In$_2$O$_3$ $\sim 3.5$ eV, indicating that in the IVO sample $E_\mathrm{F}$ is located near the bottom of the conduction band. 
We observed clear resonant enhancement in the valence band. The PES intensity as a function of photon energy ($h\nu$), i.e., constant-initial-state (CIS) spectrum, at $E_B = 1.7$ eV demonstrates that the enhancement is proportional to the intensity of the V $2p$ XAS as shown in Fig.~\ref{V3dPDOS}(a). The on- and off-resonance spectra were chosen as measured at $h\nu = 516.5$ eV and 513 eV, respectively. The difference between the on- and off-resonance spectra yields the V $3d$ PDOS. 
Although the relative intensities of different spectral features change with photon energy due to the effects of transition matrix elements and of the resonance, the structures (peak positions) of the V $3d$ PDOS are independent of photon energy, implying that the V ions are in a single electronic state. 
The V $3d$ PDOS shows a peak above the O $2p$ band, that is, in the middle of the band gap of In$_2$O$_3$, as shown in Fig.~\ref{V3dPDOS}(c). The narrow width and strong intensity of the peak imply the localized nature of the V $3d$ orbitals in the valence band, i.e., weak $\mathrm{V} \, 3d - \mathrm{O} \, 2p$ mixing in IVO.

\begin{figure}[t!]
\begin{center}
\includegraphics[width=8.8cm]{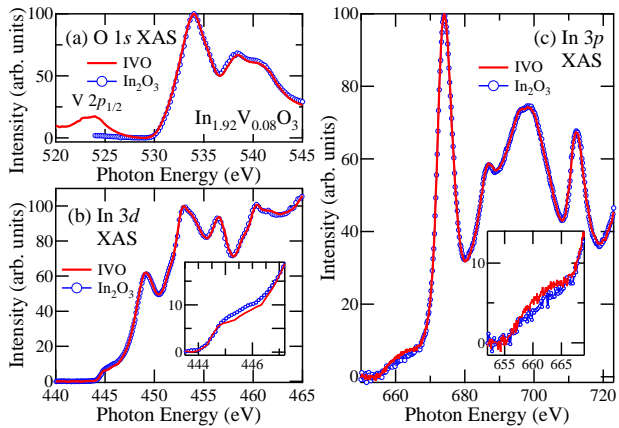}
\caption{X-ray absorption spectra of In$_{2-x}$V$_x$O$_3$ (IVO) with $x=0.08$ measured at In and O absorption edges related to the host In$_2$O$_3$. 
(a) O $K$ XAS spectra. 
(b), (c) In $3d$ and $3p$ XAS spectra, respectively. The insets show enlarged plots in the pre-edge region. 
}
\label{In2O3_XAS}
\end{center}
\end{figure}

%3-3. XAS measured at edges related to host matrix
In $n$-type DMS's, electronic states near the bottom of the conduction band are expected to be important for carrier-induced ferromagnetism since $E_\mathrm{F}$ is located there. Because the valence band of the host semiconductor is usually occupied, XAS spectra measured at the absorption edges of the host semiconductor reflect the unoccupied electronic states, namely, the conduction band, and are useful for the investigation of the electronic structure of $n$-type DMS's. 
Figure~\ref{In2O3_XAS} shows XAS spectra measured at absorption edges of In and O, and compares them between V-doped and pure In$_{2}$O$_{3}$. 
The O $1s$ XAS spectrum of IVO is nearly identical to that of In$_2$O$_3$ as shown in Fig.~\ref{In2O3_XAS}(a). 
It has been reported that the O $1s$ XAS spectrum of Fe-doped In$_2$O$_3$ differs from that of pure In$_2$O$_3$, and that the difference is induced by hybridization of the Fe $3d$ orbitals with the O $2p$ band \cite{APL_07_Jayakumar}. 
The observation implies weaker hybridization between the O $2p$ band and the V $3d$ orbitals in IVO than that between O $2p$ and Fe $3d$ in In$_{2-x}$Fe$_x$O$_3$, consistent with the result of the V $2p \to 3d$ RPES. 
In contrast, the In $3p$ and $3d$ XAS spectra were changed by V doping, i.e., there are difference between the XAS spectra of IVO and these of pure In$_{2}$O$_{3}$ in both pre-edge regions as shown in Figs.~\ref{In2O3_XAS}(b) and \ref{In2O3_XAS}(c). 
The observations clearly indicate hybridization between the In $5sp$ conduction band and the V $3d$ orbitals in IVO.

\begin{figure}[t!]
\begin{center}
\includegraphics[width=6.5cm]{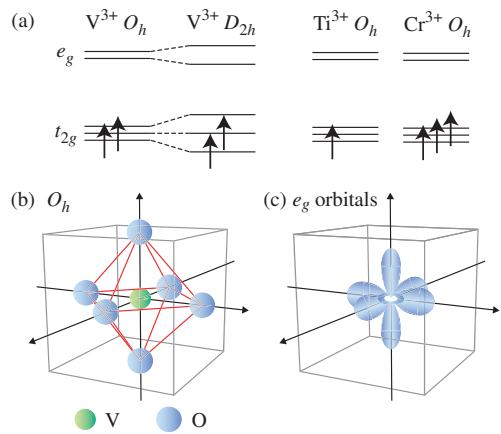}
\caption{$3d$ electronic structure of the V ion of In$_{2-x}$V$_{x}$O$_3$. 
(a) A schematic representation of the electronic structure in the V ion. Those of the Ti$^{3+}$ and Cr$^{3+}$ ion in a $O_h$ crystal field are also presented. 
(b) Octahedral coordination of the O atoms. (c) $e_g$ orbitals. 
}
\label{Electronic}
\end{center}
\end{figure}

%3-4. TM 3d-CB sp hybridization in light transition-metal doped In2O3
Based on the above findings, we shall discuss about the electronic structure of In$_2$O$_3$-based DMS's, in particular, those doped with light-TM atoms. 
Electronic structure of a TM atom octahedrally coordinated by oxygens splits into two-fold degenerate $e_g$ and three-fold degenerate $t_{2g}$ levels due to ligand crystal fields. In IVO, the V$^{3+}$ ($3d^2$) ion substituting for the In site has two electrons in the $t_{2g}$ levels (or the two lowest levels splitted from the $t_{2g}$ level due to the $D_{2h}$ symmetry) as shown in Fig.~\ref{Electronic}(a). 
Because the $e_{g}$ orbitals are directed toward the O atoms as shown in Figs.~\ref{Electronic}(b) and \ref{Electronic}(c), the O $2p$ band can hybridize with the $e_{g}$ orbitals stronger than the $t_{2g}$ orbitals. Indeed, the Slater-Koster parameter ($pd\sigma$), which represents transfer integrals between the $3d$ $e_{g}$ and ligand $p$ orbitals, has an absolute value about twice larger than ($pd\pi$), which represents transfer integrals between the $3d$ $t_{2g}$ and ligand $p$ orbitals \cite{Book_89_Harrison}. Considering the result of O $1s$ XAS, hybridization between the O $2p$ band and the $e_g$ orbitals is expected to be weak. It is probable that the $t_{2g}$ orbitals hybridize dominantly with the In band. 
Light TM ions Ti$^{3+}$ and Cr$^{3+}$ under the $O_h$ crystal field only have electron(s) in the $t_{2g}$ levels, too [Fig.~\ref{Electronic}(a)]. It follows from those arguments that for light TM-doped In$_{2}$O$_{3}$, hybridization between the host conduction band and the $3d$ $t_{2g}$ orbitals should be taken into consideration to understand their electronic properties.

%3-5. magnetic interaction
Considering the present observations, we shall discuss about the mechanism of the ferromagnetism in In$_{2-x}$V$_{x}$O. The bottom of the conduction band of In$_{2}$O$_{3}$ are mainly composed of the In $5s$ bands \cite{PRB_97_Tanaka}. For $n$-type DMS's, the electronic structure near the bottom of the conduction band is important, i.e., $s$-$d$ exchange interaction may play an essential role rather than the $p$-$d$ exchange interaction. It is possible that hybridization between the In $5s$ and V $3d$ orbitals as observed In $3p$ XAS strengthens the $s$-$d$ exchange interaction, as in the case of the host conduction band $-$ Co $3d$ orbital exchange interaction in Ti$_{1-x}$Co$_{x}$O$_2$ \cite{PRL_06_Quilty}. 
Generally speaking, because in wide-gap semiconductors the effective mass $m^\ast$ is large, the magnetic splitting due to the exchange term in Hamiltonian ($s$,$p$-$d$ exchange interaction) is greater than the spin splitting of the host valence and conduction bands predicted by ordinary $sp$ band theory, and the exchange contribution can be expressed as a large effective $g$ factor, which is given by
\begin{eqnarray*}
\label{Geff}
g_\mathrm{eff} &=& g^{\ast} + \Delta g_\mathrm{ex}, \\
\Delta g_\mathrm{ex} &=& \alpha M /\left(g_\mathrm{TM} \mu_\mathrm{B}^2H \right),
\end{eqnarray*}
where $g^\ast$ is the band $g$ factor, $g_\mathrm{TM}$ is the $g$ factor of the $3d$ ions, $\alpha$ is the exchange constant, and $\mu_\mathrm{B}$ is the Bohr magnetron, that is, $|\Delta g_\mathrm{ex}| \gg |g^{\ast}|$ for wide-gap semiconductors \cite{JAP_88_Furdyna}. 
Therefore, the $s$-$d$ exchange interaction accompanied by the spin splitting of the conduction band is expected to be dominant in In$_{2-x}$V$_{x}$O$_{3}$. In such a case, magnetic circular dichroism signal at the optical absorption edge should be measurable. Additional measurements which are related to the spin splitting of the conduction band are highly desirable. 
Since the extent of $3d$ orbitals diminish with increasing number of $3d$ electrons, effects of cation-$3d$ hybridization may weaken in heavy TM-doped DMS's.

%3-6. Oxygen vacancy
Assuming that light TM-doped In$_{2}$O$_{3}$ DMS's have common properties, we can discuss about the role of oxygen vacancies in In$_{2}$O$_{3}$-based DMS's. The oxygen vacancy is a major defect in oxides and acts as a double donor. In addition to electron doping by the vacancies, it has recently been proposed that the existence of oxygen vacancies themselves affects the ferromagnetic properties of oxide-based DMS's through the formation of the donor-impurity band (in this case, the oxygen vacancies are the centers of bound magnetic polarons) \cite{NatMater_05_Coey}, and reported that, even in non-magnetic oxide, the ferromagnetism can be induced by the oxygen vacancies \cite{PRB_06_Sundaresan, MMM_07_Hong}. 
For light TM-doped In$_{2}$O$_{3}$ DMS's, there are several works related to oxygen vacancies. Philip {\it et al}. \cite{NatMater_06_Philip} have reported that while Cr:In$_{2}$O$_{3}$ thin films deposited at high oxygen pressure (20 mTorr) show paramagnetic behavior, films deposited at low oxygen pressure (0.35 mTorr) exhibits ferromagnetic properties, where the low pressure growth induces in the thin films with electron carriers through the oxygen vacancies. Kharel {\it et al}. \cite{JAP_07_Kharel} have found that while air annealed Cr-doped In$_{2}$O$_{3}$ film and bulk samples do not show ferromagnetism, samples annealed in a high vacuum are ferromagnetic due to the creation of oxygen vacancies. 
%These reports suggest that the oxygen vacancies influence the ferromagnetism in In$_{2}$O$_{3}$-based DMS's. 
A first-principle electronic-structure calculation of In$_{2}$O$_{3}$ has predicted that when electrons are localized around the oxygen vacancy, the In-In bond is reinforced remarkably while the In-O bond is significantly weakened \cite{PRB_97_Tanaka}. Considering this result together with the experimental findings, it is possible that, besides the electron carrier doping, the presence of oxygen vacancy augments the exchange interaction between the In conduction band and the $3d$ orbitals. This effect may be stronger for light TM-doped In$_2$O$_3$ DMS's than heavy TM-doped ones because of the shrinkage of the $3d$ orbitals with the number of $3d$ electrons. 
In order to set whether this consideration is valid or not, systematic measurements of In$_{2}$O$_{3}$-based DMS's with independently controlled oxygen vacancies and carrier concentrations are desirable.

%consistent with the uniform distribution of the magnetic domains \cite{NatMater_06_Philip, JPCM_06_Hong} and the anomalous Hall effects \cite{APL_05_He, SSC_06_Kim, NatMater_06_Philip, PRB_06_Yu, PRB_07_Stankiewicz}. 

%Although both V- and Fe-doped In$_2$O$_3$ show ferromagnetic properties, it is probable that the origin of ferromagnetism differs from the kind of doped transition-metal ion if carrier-induced ferromagnetism is realized. 
%In the case of the V doping, since the conduction bands near the bottom are mainly compounded from In $5s$ band \cite{PRB_97_Tanaka}, carrier-induced ferromagnetism may be caused by Zener $s$-$d$ exchange interaction \cite{PR_51a_Zener}, which is related to spin splitting of the bottom of host conduction bands. 
%In the case of the Fe doping, the mechanism of carrier-induced ferromagnetism may be based on double-exchange \cite{PR_51b_Zener}, which is related to spin splitting of the $d$ band, or donor impurity band exchange \cite{NatMater_05_Coey}. 
%Figure~\ref{EnergyDiag} shows the energy diagram for V- and Fe-doped In$_2$O$_3$. 
%In order to investigate the origin of the ferromagnetism, systematic spectroscopic measurements such as carrier doping dependence and transition-metal concentration dependence are desirable. 

%4. CONCLUSION
In conclusion, we have performed PES and XAS studies of In$_{2-x}$V$_x$O$_3$ thin films in order to investigate their electronic structure and its relationship with ferromagnetism. From comparison with previous reports, the V $3d$ ions are expected to be trivalent states, indicating that the V ion substitutes for the In site. The V $3d$ PDOS in valence band has been observed using RPES technique and shows a sharp peak within the band gap of In$_2$O$_3$, implying the localized nature of the V $3d$ orbitals in the valence band. 
There are differences of XAS spectra measured at In $3p$ and $3d$ edges although the O $K$ XAS spectrum is independent of V doping. Based on the findings, it is likely that the $s$-$d$ exchange interaction is dominant for the ferromagnetism in In$_{2-x}$V$_{x}$O$_{3}$. 
For light TM-doped In$_{2}$O$_{3}$, the role of oxygen vacancy has been discussed. 
The present results point to a need for taking into account hybridization between the host conduction band and $3d$ orbitals in In$_{2}$O$_{3}$-based DMS's, especially for light TM doped ones. 
We believe that the findings will promote further systematic studies of $n$-type oxide-based DMS's.

%ACKNOWLEDGEMENT
This work was supported by a Grant-in-Aid for Scientific Research in Priority Area ``Spin Current: Its Creation and Control'' (19048012) from MEXT, Japan. 
Work in Sweden is supported by the Swedish Agency VINNOVA, and the Carl Tryggers Stiftelse. 
MK and MT acknowledge support from the Japan Society for the Promotion of Science for Young Scientists.

\end{document}